\documentclass[12pt]{article}
\usepackage{amsfonts}
\usepackage{amsmath}
\usepackage{amssymb}
\usepackage{axodraw}
\usepackage{mathptmx}
\usepackage[dvips]{color}
\usepackage{epsfig}
\usepackage{graphicx}
\newcommand{\calO}{{\cal O}}
\newcommand{\calP}{{\cal P}}
\newcommand{\calQ}{{\cal Q}}

\newcommand{\calY}{{\cal Y}}

\newcommand{\eV}{{\rm eV}}
\newcommand{\GeV}{{\rm GeV}}
\newcommand{\MeV}{{\rm MeV}}
\newcommand{\TeV}{{\rm TeV}}
\newcommand{\rmv}{{\rm v}}

\begin{document}
\baselineskip=16pt

\pagenumbering{arabic}

\vspace{1.0cm}

\begin{center}
{\Large\sf Neutrino mass operators of dimension up to nine in
two-Higgs-doublet model}
\\[10pt]
\vspace{.5 cm}

{Yi Liao\footnote{liaoy@nankai.edu.cn}}

{\it School of Physics, Nankai University, Tianjin 300071, China
\\
Center for High Energy Physics, Peking University, Beijing 100871,
China}

\vspace{2.0ex}

{\bf Abstract}

\end{center}

We study higher-dimensional neutrino mass operators in a low energy
theory that contains a second Higgs doublet, the two Higgs doublet
model. The operators are relevant to underlying theories in which
the lowest dimension-five mass operators would not be induced. We
list the independent operators with dimension up to nine with the
help of Young tableau. Also listed are the lowest dimension-seven
operators that involve gauge bosons and violate the lepton number by
two units. We briefly mention some of possible phenomenological
implications.

\begin{flushleft}
% corresponding to: (PACS 2010 version)
% Neutrino mass and mixing
% Ordinary neutrinos
% Extensions of electroweak Higgs sector


DOI information: 10.1016/j.physletb.2011.03.025

\end{flushleft}

\newpage

\section{Introduction}

The tiny neutrino mass can be accommodated at low energies by
nonrenormalizable, higher-dimensional mass operators. With the
lepton fields as established in the standard model (SM) and the
Higgs fields assumed to be a doublet, such operators first appear at
dimension five \cite{Weinberg:1979sa}:
\begin{eqnarray}
\calO_{xy}^{\alpha\beta}=\overline{F_{Lx}^C}\tilde H^*_\alpha\tilde
H^\dagger_\beta F_{Ly},~~~
\calP_{xy}^{\alpha\beta}=\overline{F_{Lx}^C}\tilde F_{Ly}^*\tilde
H^\dagger_\beta H_\alpha. %
\label{eq_Weinberg}
\end{eqnarray}
Here $H_\alpha$ is the $\alpha$-th Higgs doublet with hypercharge
$Y=+1$, and $F_{Lx}$ is the $x$-th left-handed leptonic doublet with
$Y=-1$. A tilde denotes the complex-conjugated field that transforms
under $SU(2)_L$ exactly as the original one, e.g., $\tilde
F_L=i\sigma^2F_L^*$, while the superscript $C$ denotes charge
conjugation with the convention $F_L^C=(F_L)^C$.

Both operators $\calO$ and $\calP$ break the lepton number by two
units. When the neutral components of the scalar doublets develop a
vacuum expectation value (VEV), $\calO$ generates a mass for
neutrinos that is inversely proportional to the energy scale
$\Lambda$ of some underlying theory responsible for the operator.
Although the operator $\calP$ does not generate a mass but involves
interactions amongst leptons and scalars of different charge, it may
arise from the same mechanism that induces $\calO$ due to the
similar structure. With a single Higgs doublet as in SM, the
operator $\calO$ is unique while $\calP$ does not exist.

It is interesting to realize that the unique operator $\calO$ in SM
may be written in three apparently different ways \cite{Ma:1998dn}.
This amounts to forming a singlet in three ways out of four factors
of the two half-isospin fields, and suggests its possible origin
from three types of seesaw mechanisms
\cite{type1,type2,Foot:1988aq}. A phenomenological issue with those
mechanisms is that the energy scale $\Lambda$ is so high that it
would not be possible to detect any other effects pertained to the
origin of neutrino mass. From the viewpoint of effective field
theory, the scale may be lowered if the mass is induced not from a
dimension-five operator but from those of even higher dimensions. It
is conceivable that there will be more and more mechanisms that can
induce a mass operator as its dimension increases, see Refs.
\cite{Babu:2009aq,Bonnet:2009ej} for some recent examples. However,
it has been established recently that the mass operator at each
higher dimension is always unique \cite{Liao:2010ku}. This implies
that as far as the neutrino mass is concerned different mechanisms
are completely equivalent. But with a lowered scale, it becomes
possible to distinguish them through other effects.

In this work we will address the neutrino mass operators in an
effective field theory that contains two Higgs doublets. Although
the two Higgs doublet model (2HDM) is interesting in itself, the
main motivation comes from supersymmetry which is a leading
candidate for physics beyond SM and is under examination at high
energy colliders. It would be also tempting to see how those
higher-dimensional operators are induced in a supersymmetric
framework. We will show that with two Higgs doublets the operators
are no more unique but increase quickly in number with their
dimension. We will list all mass operators of dimension up to nine
as well as related dimension-seven operators involving the SM gauge
fields. We will also discuss briefly some of the phenomenological
implications of these operators at low energies.

\section{Mass operators up to dimension nine}

We assume that the low energy theory contains the SM fields and an
additional Higgs doublet that also develops a VEV. The neutrinos can
only have a Majorana mass in this case. We are interested in the
high-dimensional operators that can yield a neutrino mass (called
$\calO$-type) when the Higgs fields assume their VEV's, as well as
those that do not give a mass but have a similar structure
($\calP$-type). We can therefore restrict ourselves to the
two-lepton sector that violates the lepton number by two units. The
relevant fields are the lepton doublet $F_L$, the two Higgs doublets
$H_1$ and $H_2$ plus their properly complex-conjugated fields which
also transform as a doublet under $SU(2)_L$:
\begin{eqnarray}
a=\overline{F_L^C}~(-1),~b=F_L~(-1),~c=H_1~(+1),~d=H_2~(-1),~e=\tilde
H_2~(+1),~f=\tilde H_1~(-1),
 \label{eq_notation}
\end{eqnarray}
where the number in parentheses indicates hypercharge. Our notation
is such that we always use column spinors in isospin space though
$\overline{F_L^C}$ is a row spinor in Dirac space and should appear
on the left of $F_L$ to form an appropriate Dirac bilinear. The
lepton generation index is generally inessential and can be easily
recovered when necessary. We note the following features that are
useful to exhaust all possibilities. First, since the pair $ab$
appears once, there are two more factors of $c$ or $e$ than $d$ or
$f$ to balance hypercharge. The dimension of mass operators is thus
$2n+5$, where $n$ denotes the number of copies of $d$ or $f$.
Second, the occurrence of $c$ may be replaced by $e$ if this yields
a different and nonvanishing result, and similarly with $d$ and $f$.
Finally, the SM case is recovered by the identifications $e=c$ and
$d=f$.

With an even number of fields with nonzero isospin one may imagine
to form higher isospin products before building a singlet out of
them. But this is unnecessary when all the fields are in the
fundamental representation (spinor for short) of $SU(2)$: all
isospin invariants of a given mass dimension can be exhausted by
first forming singlets from any two spinors and then multiplying
them. This is the group-theoretical reason that the three types of
seesaws reduce to the unique dimension-five Weinberg operator
$\calO$ in SM \cite{Ma:1998dn} and that its higher-dimensional
generalizations are also unique at each dimension
\cite{Liao:2010ku}.

The above point can be best seen in the tensor method in terms of
Young tableau. For $SU(2)$ a Young tableau has at most two rows, and
each column with two rows is a separate invariant. This is
especially convenient when only spinors appear, because in that case
each box represents an individual field and a two-row column is an
antisymmetric, invariant product of the two spinors involved. This
has a few immediate consequences. First, there can be no bare mass
term from $\overline{F_L^C}F_L$ even if $F_L$ had a zero
hypercharge. Second, denoting a spinor by its index in the box, we
have the basic relation:
%\begin{center}
\begin{eqnarray}
\begin{picture}(150,20)(0,0)

\SetOffset(0,10)%
\Line(20,-10)(0,-10)\Line(0,0)(20,0)\Line(20,10)(0,10)
\Line(0,-10)(0,10)\Line(10,-10)(10,10)\Line(20,-10)(20,10)
\Text(5,5)[]{$i$}\Text(15,5)[]{$m$}
\Text(5,-5)[]{$j$}\Text(15,-5)[]{$n$}%
\Text(35,0)[]{$-$}

\SetOffset(50,10)%
\Line(20,-10)(0,-10)\Line(0,0)(20,0)\Line(20,10)(0,10)
\Line(0,-10)(0,10)\Line(10,-10)(10,10)\Line(20,-10)(20,10)
\Text(5,5)[]{$i$}\Text(15,5)[]{$m$}
\Text(5,-5)[]{$n$}\Text(15,-5)[]{$j$}%
\Text(35,0)[]{$=$}

\SetOffset(100,10)%
\Line(20,-10)(0,-10)\Line(0,0)(20,0)\Line(20,10)(0,10)
\Line(0,-10)(0,10)\Line(10,-10)(10,10)\Line(20,-10)(20,10)
\Text(5,5)[]{$i$}\Text(15,5)[]{$j$}
\Text(5,-5)[]{$m$}\Text(15,-5)[]{$n$}

\end{picture}
\label{eq_basic}
\end{eqnarray}
%\end{center}
which is equivalent to the relation ($i,j,m,n=1,2$)
\begin{eqnarray}
\epsilon_{ij}\epsilon_{mn}-\epsilon_{in}\epsilon_{mj}
=\epsilon_{im}\epsilon_{jn}.
\end{eqnarray}
Applied to the dimension-five Weinberg operators in eq
(\ref{eq_Weinberg}), we have
\begin{eqnarray}
\calO_{xy}^{\alpha\beta}-\calO_{yx}^{\alpha\beta}
=\calP_{xy}^{\alpha\beta},
\end{eqnarray}
which means that only one group of dimension-five operators (type
$\calO$) listed in Ref \cite{Weinberg:1979sa} are actually
independent. (Be careful not to mix the generation indices with the
spinor indices.) More generally, putting spinors directly in boxes
we have
%\begin{center}
\begin{eqnarray}
\begin{picture}(300,20)(0,0)

\SetOffset(70,10)%
\Line(40,-10)(0,-10)\Line(0,0)(20,0)\Line(40,10)(0,10)
\Line(0,-10)(0,10)\Line(10,-10)(10,10)\Line(20,-10)(20,10)
\Line(40,-10)(40,10)%
\Text(5,5)[]{$a$}\Text(5,-5)[]{$b$}
\Text(15,5)[]{$\kappa$}\Text(15,-5)[]{$\sigma$}\Text(30,0)[]{$\calY$}
\Text(50,0)[]{$=$}

\SetOffset(130,10)%
\Line(40,-10)(0,-10)\Line(0,0)(20,0)\Line(40,10)(0,10)
\Line(0,-10)(0,10)\Line(10,-10)(10,10)\Line(20,-10)(20,10)
\Line(40,-10)(40,10)%
\Text(5,5)[]{$a$}\Text(15,5)[]{$b$}
\Text(5,-5)[]{$\kappa$}\Text(15,-5)[]{$\sigma$}\Text(30,0)[]{$\calY$}
\Text(50,0)[]{$-$}

\SetOffset(190,10)%
\Line(40,-10)(0,-10)\Line(0,0)(20,0)\Line(40,10)(0,10)
\Line(0,-10)(0,10)\Line(10,-10)(10,10)\Line(20,-10)(20,10)
\Line(40,-10)(40,10)%
\Text(5,5)[]{$a$}\Text(15,5)[]{$b$}
\Text(5,-5)[]{$\sigma$}\Text(15,-5)[]{$\kappa$}\Text(30,0)[]{$\calY$}

\end{picture}
\end{eqnarray}
%\end{center}
where $\calY$ is any Young tableau. Namely, the $\calP$-type
operators that contain as a factor an invariant formed out of $a,~b$
are linear compositions of the $\calO$-type operators. By making a
complete list of all mass operators (of type-$\calO$), all non-mass
operators (of type-$\calP$) with a similar structure are
automatically covered. In the language of Young tableau, we will
never put $a,~b$ in the same column.

It is easy to figure out all dimension-five operators since $d,~f$
cannot appear while $c/e$ appears twice. They are
\begin{eqnarray}
\begin{picture}(150,20)(0,0)

\SetOffset(0,10)%
\Line(20,-10)(0,-10)\Line(0,0)(20,0)\Line(20,10)(0,10)
\Line(0,-10)(0,10)\Line(10,-10)(10,10)\Line(20,-10)(20,10)
\Text(5,5)[]{$a$}\Text(15,5)[]{$b$}
\Text(5,-5)[]{$c$}\Text(15,-5)[]{$c$}

\SetOffset(40,10)%
\Line(20,-10)(0,-10)\Line(0,0)(20,0)\Line(20,10)(0,10)
\Line(0,-10)(0,10)\Line(10,-10)(10,10)\Line(20,-10)(20,10)
\Text(5,5)[]{$a$}\Text(15,5)[]{$b$}
\Text(5,-5)[]{$c$}\Text(15,-5)[]{$e$}

\end{picture}
\end{eqnarray}
plus those obtained by $c\leftrightarrow e$, or
\begin{eqnarray}
&&S_5=(a,c)_0(b,c)_0,~T_5=(a,c)_0(b,e)_0,\nonumber\\
&&\bar S_5=S_5|_{c\leftrightarrow e},~\bar
T_5=T_5|_{c\leftrightarrow e},
\end{eqnarray}
where the subscript $0$ denotes an isospin invariant formed by
antisymmetrizing the fields inside the parentheses; for instance,
denoting the upper (lower) component of a spinor by a subscript plus
(minus) sign, we have $\sqrt{2}(a,c)_0=a_+c_--a_-c_+$. Since $a$ and
$b$ are essentially the same field, the list of operators may be
further reduced. To see this clearly, we reserve the lepton
generation index by putting $a=\overline{F_{Lx}^C}$, $b=F_{Ly}$.
Then,
\begin{eqnarray}
2\bar T_5^{xy}=
\big(\overline{\nu_{Lx}^C}e_--\overline{f_{Lx}^C}e_+\big)
\big(\nu_{Ly}c_--f_{Ly}c_+\big)=
\big(\overline{\nu_{Ly}^C}c_--\overline{f_{Ly}^C}c_+\big)
\big(\nu_{Lx}e_--f_{Lx}e_+\big)=2T_5^{yx},
\end{eqnarray}
where $\overline{\psi_i^C}\psi_j=\overline{\psi_j^C}\psi_i$ is used.
We can thus choose $S_5,~\bar S_5,~T_5$ as the complete and
independent list of dimension-five operators.

At dimension seven, the operators contain three copies of $c$ or $e$
and one copy of $d$ or $f$, and can be classified as $S:~c^3d$,
$T:~c^2ed$, plus those obtained by $c\leftrightarrow e$, or
$d\leftrightarrow f$, or both interchanges. The first one is easy to
write down:
\begin{eqnarray}
S_7=(a,c)_0(b,c)_0(d,c)_0.
\end{eqnarray}
For the second one, there are following possibilities to distribute
the spinors in the boxes of a $2\times 3$ Young tableau:
%\begin{center}
\begin{eqnarray}
\begin{picture}(300,20)(0,0)

\SetOffset(0,10)%
\Line(30,-10)(0,-10)\Line(0,0)(30,0)\Line(30,10)(0,10)
\Line(0,-10)(0,10)\Line(10,-10)(10,10)\Line(20,-10)(20,10)
\Line(30,-10)(30,10)%
\Text(5,5)[]{$a$}\Text(15,5)[]{$b$}\Text(25,5)[]{\Red{$d$}}
\Text(5,-5)[]{\Blue{$c$}}\Text(15,-5)[]{\Blue{$c$}}
\Text(25,-5)[]{\Red{$e$}}%
\Text(15,-15)[]{1st}

\SetOffset(50,10)%
\Line(30,-10)(0,-10)\Line(0,0)(30,0)\Line(30,10)(0,10)
\Line(0,-10)(0,10)\Line(10,-10)(10,10)\Line(20,-10)(20,10)
\Line(30,-10)(30,10)%
\Text(5,5)[]{$a$}\Text(15,5)[]{$b$}\Text(25,5)[]{\Red{$d$}}
\Text(5,-5)[]{\Blue{$c$}}\Text(15,-5)[]{\Red{$e$}}
\Text(25,-5)[]{\Blue{$c$}}%
\Text(15,-15)[]{2nd}

\SetOffset(100,10)%
\Line(30,-10)(0,-10)\Line(0,0)(30,0)\Line(30,10)(0,10)
\Line(0,-10)(0,10)\Line(10,-10)(10,10)\Line(20,-10)(20,10)
\Line(30,-10)(30,10)%
\Text(5,5)[]{$a$}\Text(15,5)[]{$b$}\Text(25,5)[]{\Red{$e$}}
\Text(5,-5)[]{\Blue{$c$}}\Text(15,-5)[]{\Red{$d$}}
\Text(25,-5)[]{\Blue{$c$}}%
\Text(15,-15)[]{3rd}

\SetOffset(150,10)%
\Line(30,-10)(0,-10)\Line(0,0)(30,0)\Line(30,10)(0,10)
\Line(0,-10)(0,10)\Line(10,-10)(10,10)\Line(20,-10)(20,10)
\Line(30,-10)(30,10)%
\Text(5,5)[]{$a$}\Text(15,5)[]{$b$}\Text(25,5)[]{\Red{$d$}}
\Text(5,-5)[]{\Red{$e$}}\Text(15,-5)[]{\Blue{$c$}}
\Text(25,-5)[]{\Blue{$c$}}%
\Text(15,-15)[]{4th}

\SetOffset(200,10)%
\Line(30,-10)(0,-10)\Line(0,0)(30,0)\Line(30,10)(0,10)
\Line(0,-10)(0,10)\Line(10,-10)(10,10)\Line(20,-10)(20,10)
\Line(30,-10)(30,10)%
\Text(5,5)[]{$a$}\Text(15,5)[]{$b$}\Text(25,5)[]{\Red{$e$}}
\Text(5,-5)[]{\Red{$d$}}\Text(15,-5)[]{\Blue{$c$}}
\Text(25,-5)[]{\Blue{$c$}}%
\Text(15,-15)[]{5th}

\end{picture}
\label{eq_Young2}
\end{eqnarray}
%\end{center}
But the basic relation in eq (\ref{eq_basic}) implies
\begin{eqnarray}
\textrm{1st}-\textrm{2nd}+\textrm{3rd}=0,~
\textrm{1st}-\textrm{4th}+\textrm{5th}=0,
\end{eqnarray}
which eliminate two operators. We choose the 1st, 3rd, and 5th ones
to be independent:
\begin{eqnarray}
T_7^1=(a,c)_0(b,c)_0(d,e)_0,~T_7^2=(a,c)_0(b,d)_0(e,c)_0,~
T_7^3=(a,d)_0(b,c)_0(e,c)_0.
\end{eqnarray}
But for the same reason as for $\bar T_5$, $T_7^3$ is covered by
$T_7^2$ when the lepton generation indices are reserved, and may
thus be excluded as redundant. The remaining operators are obtained
by interchanges:
\begin{eqnarray}
&&\bar S_7=S_7|_{c\leftrightarrow e},~\hat
S_7=S_7|_{d\leftrightarrow f},~\tilde S_7=S_7|_{c\leftrightarrow
e,d\leftrightarrow f};\nonumber\\
&&\bar T_7^{1,2}=T_7^{1,2}|_{c\leftrightarrow e},~\hat
T_7^{1,2}=T_7^{1,2}|_{d\leftrightarrow f},~ \tilde
T_7^{1,2}=T_7^{1,2}|_{c\leftrightarrow e,d\leftrightarrow f}.
\end{eqnarray}
There are altogether 12 operators at dimension seven.

The dimension-nine operators contain four copies of $c$ or $e$ and
two copies of $d$ or $f$, which are classified as
$S:~c^4d^2,~T:~c^4df,~U:~c^3ed^2,~V:~c^3edf,~W:~c^2e^2d^2,~
X:~c^2e^2df$, plus those obtained by interchange $c\leftrightarrow
e$, or $d\leftrightarrow f$, or both. We continue to denote an
operator obtained by $c\leftrightarrow e$ with a bar, that by
$d\leftrightarrow f$ with a hat, and the one by both
$c\leftrightarrow e$ and $d\leftrightarrow f$ with a tilde. It is
easy to write down $S$ and $T$:
\begin{eqnarray}
S_9=S_7(d,c)_0,~T_9=\hat S_7(d,c)_0.
\end{eqnarray}
And there are four more operators that are obtained by interchanges:
\begin{eqnarray}
\bar S_9,~\hat S_9,~\tilde S_9;~\bar T_9.
\end{eqnarray}
The $U$ operators have one more factor of $cd$ than $T_7$. Deleting
the redundant one associated with $T_7^3$, we have
\begin{eqnarray}
U_9^{1,2}=T_7^{1,2}(d,c)_0.
\end{eqnarray}
The additional operators obtained by interchanges are also
independent
\begin{eqnarray}
\bar U_9^{1,2},~\hat U_9^{1,2},~\tilde U_9^{1,2}.
\end{eqnarray}

It is possible to distribute in nine ways the spinors of $V$ in a
$2\times 4$ Young tableau, but only four of them yield independent
operators. Five Young tableaux are obtained from those in eq
(\ref{eq_Young2}) by attaching an additional column of $(f,c)_0$,
thus giving the three independent operators:
\begin{eqnarray}
V_9^{1,2,3}=T_7^{1,2,3}(f,c)_0.
\end{eqnarray}
Another three tableaux are obtained from the above by
$d\leftrightarrow f$: $\hat V_9^{1,2,3}$, and the ninth one
corresponds to
\begin{eqnarray}
V_9^0=(a,c)_0(b,c)_0(d,f)_0(e,c)_0.
\end{eqnarray}
Similar to $T_5^3$ and $T_7^3$, $V_9^3$ can be deleted from the
list. Furthermore, the basic relation (\ref{eq_basic}) implies that
\begin{eqnarray}
\hat V_9^1-V_9^1=V_9^2-\hat V_9^2=V_9^3-\hat V_9^3=-V_9^0,
\end{eqnarray}
so that we can keep $V_9^{0,1,2}$ in the list while excluding $\hat
V_9^{1,2,3}$ as redundant. Finally, there are three more operators
obtained from interchange $c\leftrightarrow e$:
\begin{eqnarray}
\bar V_9^{0,1,2}.
\end{eqnarray}
A similar (but slightly different) analysis applies to the operators
$W$, which have four independent forms
\begin{eqnarray}
W_9^{1,2}=T_7^{1,2}(d,e)_0,~\bar W_9^{1,2},
\end{eqnarray}
plus four more by interchange $d\leftrightarrow f$:
\begin{eqnarray}
\hat W_9^{1,2},~\tilde W_9^{1,2}.
\end{eqnarray}

Finally we come to the symmetric case of $X$ that has the most
possible Young tableaux (18 in total). The basic relation
(\ref{eq_basic}) removes ten of them as redundant and the symmetry
in the lepton fields deletes another three, leaving us with five
independent operators:
\begin{eqnarray}
&&X_9^{1,2}=T_7^{1,2}(f,e)_0,~\bar X_9^1\nonumber\\
&&X_9^A=(a,c)_0(b,e)_0(d,f)_0(c,e)_0,\nonumber\\
&&X_9^S=(a,d)_0(b,f)_0(c,e)_0(c,e)_0,
\end{eqnarray}
where $X_9^S$ ($X_9^A$) is (anti)symmetric in $c\leftrightarrow e$
and $d\leftrightarrow f$ respectively when the lepton generation
indices are ignored. There are altogether 33 dimension-nine mass
operators.

\section{Adding gauge bosons}

The underlying physics that produces the higher-dimensional neutrino
mass operators in the last section may also induce lepton-number
violating interactions with gauge bosons. In this section we
continue to work in the two-lepton sector and list the lowest
dimension-seven operators with gauge bosons that are built upon the
dimension-five mass operators. The gauge fields may enter in two
ways, either through gauge covariant derivatives or through field
strength tensors. The first case amounts to introducing new Lorentz
vector fields that have the same quantum numbers under the SM gauge
group as the original fields, $a,~b,~c,~e$. The second case requires
that those original fields must be built into a hypercharge-neutral,
isospin-triplet or -singlet form that couples to the field strength
tensors of $SU(2)_L$ and $U(1)_Y$ respectively.

We start with the operators containing the gauge covariant
derivative
\begin{eqnarray}
D_\mu=\partial_\mu-ig_2\frac{1}{2}\sigma^aW^a_\mu\mp
ig_1\frac{1}{2}B_\mu,
\end{eqnarray}
where the minus (plus) sign applies to the fields $c,~e$ ($a,~b$),
and $W^a_\mu$ and $B_\mu$ are the gauge fields with gauge couplings
$g_{2,1}$. Distributing two factors of $D_\mu$ to any two of the
four fields in the mass operators $S_5$ and $T_5$ yields
\begin{eqnarray}
J^{1,\cdots,6}&=&(D_\mu a,D^\mu c)_0(b,c)_0,~(D_\mu a,c)_0(D^\mu
b,c)_0,~ (D_\mu a,c)_0(b,D^\mu c)_0,\nonumber\\
&&(a,D_\mu c)_0(b,D^\mu c)_0,~(a,D_\mu c)_0(D^\mu b,c)_0,
~(a,c)_0(D_\mu b,D^\mu c)_0;\\
K^{1,\cdots,6}&=&(D_\mu a,D^\mu c)_0(b,e)_0,~(D_\mu a,c)_0(D^\mu
b,e)_0,~ (D_\mu a,c)_0(b,D^\mu e)_0,\nonumber\\
&&(a,D_\mu c)_0(b,D^\mu e)_0,~(a,D_\mu c)_0(D^\mu b,e)_0,~
(a,c)_0(D_\mu b,D^\mu e)_0;
\end{eqnarray}
plus $\bar J^{1,\cdots,6}$ and $\bar K^{1,\cdots,6}$ that are
obtained by $c\leftrightarrow e$. Since the gauge covariant
derivative does not spoil the relation $\overline{\psi_i^C}\psi_j=
\overline{\psi_j^C}\psi_i$, we can exclude some of the operators as
redundant as we did with $\bar T_5$. Reserving the lepton generation
indices and denoting the upper (lower) component of a gauge
covariant derivative also by a subscript plus (minus) sign, we have,
for instance, $a_{x+}(D_\mu b_y)_-=(D_\mu a_y)_-b_{x+}$ using our
notations in eq (\ref{eq_notation}). It should be reminded that no
integration by parts can be legitimately used here; instead, the
relation $\overline{\psi_i^C}\psi_j= \overline{\psi_j^C}\psi_i$ is
sufficient. Some inspection then shows that $J_{xy}^6=J_{yx}^1$,
$J_{xy}^5=J_{yx}^3$ and similarly for $\bar J$. Since the
$K$operators involve simultaneously $c$ and $e$ fields, a stronger
reduction of operators becomes possible, namely, $\bar
K_{xy}^{1,2,3,4,5,6}=K_{yx}^{6,2,5,4,3,1}$. The complete and
independent operators can thus be chosen to be
\begin{eqnarray}
J^{1,\cdots,4},~\bar J^{1,\cdots,4},~K^{1,\cdots,6}.
\end{eqnarray}

To construct dimension-seven operators involving gauge field
strength tensors, the Lorentz indices of the tensors must be
contracted by Dirac matrices. This means that a $\sigma^{\mu\nu}$
should be sandwiched between the lepton fields $a$ and $b$, which
fits well with their chiralities. Consider first the coupling of
$B_{\mu\nu}$ to a singlet formed from $abce$. The only difference to
the dimension-five mass operator $T_5$ is to insert a
$\sigma^{\mu\nu}$ between $a$ and $b$:
\begin{eqnarray}
M(B)=(a,c)_0\sigma^{\mu\nu}(b,e)_0B_{\mu\nu}.
\end{eqnarray}
Note that $\bar M(B)$, which is again obtained from $M(B)$ by the
interchange $c\leftrightarrow e$, is not independent since when
attaching the lepton generation indices to $a$ and $b$ we have $\bar
M_{xy}(B)=-M_{yx}(B)$ upon using
$\overline{\psi^C_x}\sigma_{\mu\nu}\psi_y=-\overline{\psi_y^C}
\sigma_{\mu\nu}\psi_x$. It is not necessary either to consider the
case where $a,~b$ lie in the same column of a tableau since the
basic relation (\ref{eq_basic}) is not disturbed by the Lorentz
structure. Similarly, the counterparts of $S_5$ and $\bar S_5$ are
\begin{eqnarray}
L(B)=(a,c)_0\sigma^{\mu\nu}(b,c)_0B_{\mu\nu},~\bar L(B).
\end{eqnarray}

In contrast to the above, the $SU(2)_L$ gauge field strength
$W^a_{\mu\nu}$ being a triplet must couple to the triplet states of
$abce$ to become a singlet. There are apparently nine possibilities
for $abce$ to form a triplet state: one pair of spinors in a singlet
and the other in a triplet (six in total), or both pairs in a
triplet multiplied into a triplet (three in total). But only three
of them are independent as we show below. We note first of all that
there are four possible ways to form a state with $I_3=+1$, which do
not necessarily have a definite total isospin $I$:
\begin{eqnarray}
w=a_-b_+c_+e_+,~x=a_+b_-c_+e_+,~ y=a_+b_+c_-e_+,~z=a_+b_+c_+e_-.
\end{eqnarray}
But symmetry requires that the $I_3=+1$ state of a triplet ($I=1$)
formed from four spinors be a difference of the above quantities,
and thus there can only be three independent states with $I_3=+1$
belonging to three triplets. (The fourth $I_3=+1$ state has $I=2$,
and all this is consistent with isospin composition indeed.) For
instance, the $I_3=+1$ state formed from $ab$ in a singlet and $ce$
in a triplet is $(w-x)/\sqrt{2}$, while the $I_3=+1$ state with all
of $ab$, $ce$, and $abce$ in a triplet is given by $(w+x-y-z)/2$.

To write isospin-1 states formed with two isospin-half ones, it is
convenient to use the row spinor. We denote by a check the combined
action of tilde and dagger on the isospin space, which transforms a
column spinor to a row spinor in the complex representation. For
instance,
\begin{eqnarray}
\check b=\tilde b^\dagger=(f_L,-\nu_L),
\end{eqnarray}
can form a singlet with $c$ and the gauge field strength, $\check b
W_{\mu\nu}c$, where
$W_{\mu\nu}\equiv\frac{1}{2}\sigma^aW_{\mu\nu}^a$. The complete and
independent couplings of $W_{\mu\nu}^a$ to $abce$ are therefore as
follows:
\begin{eqnarray}
M^1(W)=(a,\sigma^{\mu\nu}b)_0\check c W_{\mu\nu}e,~
M^2(W)=(a,c)_0\sigma^{\mu\nu}\check b W_{\mu\nu}e,~\bar M^2(W).
\end{eqnarray}
$\bar M^2(W)$ is independent of $M^2(W)$ since $c$ and $e$ are now
at inequivalent places in contrast to $M(B)$. On the other hand,
$\bar M^1(W)=M^1(W)$ because $\check c W_{\mu\nu}e=\check e
W_{\mu\nu}c$. This is in accord with the above symmetry arguments.
The operators with two $c$ or two $e$ are
\begin{eqnarray}
L^1(W)=(a,\sigma^{\mu\nu}b)_0\check c W_{\mu\nu}c,~
L^2(W)=(a,c)_0\sigma^{\mu\nu}\check b W_{\mu\nu}c,~\bar L^1(W),~\bar
L^2(W).
\end{eqnarray}
To summarize, the complete and independent dimension-seven operators
involving the gauge field tensors are $L(B),~\bar L(B)$, $M(B)$,
$L^1(W),~L^2(W)$, $\bar L^1(W),~\bar L^2(W)$, $M^1(W),~M^2(W)$ and
$\bar M^2(W)$.

\section{Discussion}

The effective operators that we have written down in the last two
sections involve various lepton-number violating interactions of
multi-Higgs and gauge bosons, which may have rich phenomenological
implications. But to make a complete analysis, we should include
some other operators at a similar dimension, in particular those
involving four-fermions, that violate the lepton number by two
units. Operators involving four and six fermions in SM were analyzed
in Refs. \cite{Babu:2001ex,de Gouvea:2007xp} for inducing neutrino
mass at the loop level and their phenomenology explored in \cite{de
Gouvea:2007xp}. The neutrino mass operators with two Higgs doublets
were symbolically written down in Ref. \cite{Bonnet:2009ej} from
hypercharge balance but no attempt was made to complete their
isospin structures. Instead, possible underlying models were
suggested that could induce a specific dimension-seven operator via
seesaw, together with radiative mechanisms.

In this concluding section, we discuss briefly some interesting
interactions contained in the operators listed in the last sections,
while leaving a more complete phenomenological analysis for the
future work. Assume both Higgs doublets develop VEV's which are
generally complex with phases $u_{1,2}$,
\begin{eqnarray}
\langle H_1^0\rangle=\frac{\rmv}{\sqrt{2}}u_1c_\beta,~ \langle
H_2^0\rangle=\frac{\rmv}{\sqrt{2}}u_2s_\beta,
\end{eqnarray}
where $\rmv=246~\GeV$ and $c_\beta=\cos\beta,~s_\beta=\sin\beta$.
The would-be Goldstone bosons $G^{\pm,0}$ and physical scalars
$H^\pm,~A^0,~R_\alpha$ ($\alpha=1,~2$) are related to the original
fields by unitary transformations:
\begin{eqnarray}
\left(\begin{array}{c}G^-\\H^-\end{array}\right)
&=&\left(\begin{array}{cc}u_2^*s_\beta&-u_1c_\beta\\
u_1^*c_\beta&u_2s_\beta\end{array}\right)
\left(\begin{array}{c}H_2^-\\H_1^-\end{array}\right),\\
\left(\begin{array}{c}iI_\alpha\\R_\alpha\end{array}\right)
&=&\frac{1}{\sqrt{2}}\left(\begin{array}{cc}
u_\alpha^*&-u_\alpha\\u_\alpha^*&u_\alpha\end{array}\right)
\left(\begin{array}{c}H_\alpha^0\\H_\alpha^{0*}\end{array}\right),
\\
\left(\begin{array}{c}G^0\\A^0\end{array}\right)
&=&\left(\begin{array}{cc}s_\beta&-c_\beta\\
c_\beta&s_\beta\end{array}\right)
\left(\begin{array}{c}I_2\\I_1\end{array}\right).
\end{eqnarray}
When CP is conserved, $A^0$ is a pseudoscalar while $R_{1,2}$ are
scalars whose mixing is determined by the scalar potential.

Attaching the lepton generation indices, the operator $T_5$, for
instance, contains a term
\begin{eqnarray}
-\frac{1}{4}u_1u_2^*\rmv^2c_\beta s_\beta
\overline{\nu_{Lx}^C}\nu_{Ly},
\end{eqnarray}
which gives neutrino mass after incorporating a coefficient matrix
in generations. The phases $u_1u_2^*$ can be removed by redefinition
of fields, but will reappear in other terms of $T^5$ that involve
the Higgs scalars and a lepton pair. These interactions are
relatively hard to explore since the dominant decays of the scalars
generically depend on the details of the underlying theory.
Furthermore, as we discussed in Introduction, to have any chance at
all to discover the mass generation mechanism, the mass should be
generated from operators of a high enough dimension so that the
relevant physics scale could be lowered. A promising scenario would
be that the mass operators are generated at, say, dimension nine,
while the lepton-number violating operators involving gauge fields
are generated at dimension seven by the same physics through
tree-level or one-loop effects. For these interactions we can say
something more certain since we know how the gauge bosons interact
with the SM fermions. We therefore will concentrate on them in what
follows.

Consider first the operators involving gauge field tensors, for
instance, $M(B)$. In addition to terms involving scalars, it
contains the dipole interactions with the $Z$ boson and photon $A$:
\begin{eqnarray}
+\frac{1}{4}u_1u_2^*\rmv^2c_\beta s_\beta
(s_WZ_{\mu\nu}-c_WA_{\mu\nu})\overline{\nu_{Lx}^C}
\sigma^{\mu\nu}\nu_{Ly},
\end{eqnarray}
with $c_W=\cos\theta_W$ and $s_W=\sin\theta_W$. A similar dipole
term also appears in $M^2(W)$:
\begin{eqnarray}
-\frac{1}{4\sqrt{2}}u_1u_2^*\rmv^2c_\beta s_\beta
(c_WZ_{\mu\nu}+s_WA_{\mu\nu})\overline{\nu_{Lx}^C}\sigma^{\mu\nu}\nu_{Ly}.
\end{eqnarray}
These terms are also contained in the corresponding $L$ operators
and barred operators except for different factors of $u_{1,2}$,
$c_\beta$ and $s_\beta$. For any given value of $\beta$ there are
always operators that are not suppressed by its triangular
functions. We assign a common coefficient $eC_\textrm{d.m.}$ to (the
sum) of those operators while ignoring factors of order one. While
Majorana neutrinos have no dipole moments due to CPT invariance,
they can accommodate transition dipole moments between different
neutrinos \cite{Kayser:1982br}. Roughly speaking, the upper bounds
on the latter are about $10^{-10}\mu_B$ or weaker from laboratory
experiments \cite{Auerbach:2001wg} and about $10^{-12}\mu_B$ from
astrophysical arguments on energy loss in stars
\cite{Raffelt:1999gv}. Here $\mu_B=e/(2m_e)$ is the Bohr magneton.
They translate into a bound on the coefficient of the operators:
\begin{eqnarray}
C_\textrm{d.m.}\lesssim\frac{10^{-10}\textrm{ or
}10^{-12}}{m_e\rmv^2},\textrm{ i.e., }
C_\textrm{d.m.}\lesssim(6.7\textrm{ or }31~\TeV)^{-3}.
\end{eqnarray}

For the operators involving gauge covariant derivatives, the most
interesting interaction is the one that contributes to the
neutrinoless double beta decay,
\begin{eqnarray}
&&-J^{2,4}=J^{3}=\frac{1}{2}m_W^2u_1^2c_\beta^2\calQ_{xy}+\cdots,
\nonumber\\
&&-\bar J^{2,4}=\bar J^{3}
=\frac{1}{2}m_W^2(u_2^*)^2s_\beta^2\calQ_{xy}+\cdots,
\nonumber\\
&&K^{2,4}=-K^{3,5}=\frac{1}{2}m_W^2u_1u_2^*c_\beta
s_\beta\calQ_{xy}+\cdots,
\end{eqnarray}
while $J^{1,6},~K^{1,6},~\bar J^{1,6}$ do not contain the operator
$\calQ_{xy}=\overline{f_{Lx}^C}f_{Ly}W^{+\mu}W_\mu^+$. Here
$m_W=\frac{1}{2}g_2\rmv$ is the $W^\pm$ boson mass. We assign a
common coefficient $C_{xy}$ to (the sum) of these operators. Barring
exceptional cancellation, their contribution to the subprocess
$W^-W^-\to ee$, $\sim C_{ee}m_W^2$, should not exceed the usual one
via the exchange of light active neutrinos, which is experimentally
bounded and given by $\sim m_{ee}/q^2$. Here
$m_{ee}=\sum_jm_jU^2_{ej}$ with $m_j$ being the mass of the neutrino
$\nu_j$ and $U$ the leptonic mixing matrix, and $q\sim (50\sim
100)~\MeV$ is the momentum transfer. The upper bounds on $m_{ee}$
\cite{KlapdorKleingrothaus:2004ge} then imply that
\begin{eqnarray}
|C_{ee}|\lesssim\frac{|m_{ee}|}{|q^2|m_W^2}\sim(5~\TeV)^{-3},
\end{eqnarray}
where we assume for order of magnitude estimation, $|m_{ee}|\sim
0.5~\eV$ and $q\sim 100~\MeV$.

The operators displayed in the last section also contain other
interactions involving multiple scalars and gauge bosons, or modify
the SM interactions. We leave this more complete phenomenological
analysis for the future work which should better include the effects
of multiple-fermion operators with a comparable dimension.

%\newpage
\vspace{0.5cm}
\noindent %
{\bf Acknowledgement}

This work is supported in part by the grants NSFC-10775074,
NSFC-10975078, NSFC-11025525, and the 973 Program 2010CB833000.

%\vspace{0.5cm}
\noindent %
%\newpage
%\baselineskip=20pt


\begin{thebibliography}{100}
%\begin{enumerate}

%\cite{Weinberg:1979sa}
\bibitem{Weinberg:1979sa}
  S.~Weinberg,
  %``Baryon And Lepton Nonconserving Processes,''
  Phys.\ Rev.\ Lett.\  {\bf 43}, 1566 (1979).
  %%CITATION = PRLTA,43,1566;%%

%\cite{Ma:1998dn}
\bibitem{Ma:1998dn}
  E.~Ma,
  %``Pathways to Naturally Small Neutrino Masses,''
  Phys.\ Rev.\ Lett.\  {\bf 81}, 1171 (1998)
  [arXiv:hep-ph/9805219].
  %%CITATION = PRLTA,81,1171;%%

\bibitem{type1}M. Gell-Mann, P. Ramond, R. Slansky, in:
D. Freedman, P. van Nieuwenhuizen (Eds.), Supergravity,
North-Holland, Amsterdam, 1979, p.315; T. Yanagida, in: O. Sawada,
A. Sugamoto (Eds.), Proceedings of the Workshop on Unified Theory
and Baryon Number in the Universe, KEK, Japan, 1979; R.N. Mohapatra,
G. Senjanovic, Phys. Rev. Lett. {\bf 44} (1980) 912.

\bibitem{type2}
%\cite{Konetschny:1977bn}
%\bibitem{Konetschny:1977bn}
  W.~Konetschny and W.~Kummer,
  %``Nonconservation Of Total Lepton Number With Scalar Bosons,''
  Phys.\ Lett.\  B {\bf 70}, 433 (1977);
  %%CITATION = PHLTA,B70,433;%%
%\cite{Cheng:1980qt}
%\bibitem{Cheng:1980qt}
  T.~P.~Cheng and L.~F.~Li,
  %``Neutrino Masses, Mixings And Oscillations In SU(2) X U(1) Models Of
  %Electroweak Interactions,''
  Phys.\ Rev.\  D {\bf 22}, 2860 (1980);
  %%CITATION = PHRVA,D22,2860;%%
%\cite{Schechter:1980gr}
%\bibitem{Schechter:1980gr}
  J.~Schechter and J.~W.~F.~Valle,
  %``Neutrino Masses In SU(2) X U(1) Theories,''
  Phys.\ Rev.\  D {\bf 22}, 2227 (1980).
  %%CITATION = PHRVA,D22,2227;%%

%\cite{Foot:1988aq}
\bibitem{Foot:1988aq}
  R.~Foot, H.~Lew, X.~G.~He and G.~C.~Joshi,
  %``SEESAW NEUTRINO MASSES INDUCED BY A TRIPLET OF LEPTONS,''
  Z.\ Phys.\  C {\bf 44}, 441 (1989).
  %%CITATION = ZEPYA,C44,441;%%

%\cite{Babu:2009aq}
\bibitem{Babu:2009aq}
  K.~S.~Babu, S.~Nandi and Z.~Tavartkiladze,
  %``New Mechanism for Neutrino Mass Generation and Triply Charged Higgs Bosons
  %at the LHC,''
  Phys.\ Rev.\  D {\bf 80}, 071702 (2009)
  [arXiv:0905.2710 [hep-ph]];
  %%CITATION = PHRVA,D80,071702;%%
%\cite{Xing:2009hx}
%\bibitem{Xing:2009hx}
  Z.~z.~Xing and S.~Zhou,
  %``Multiple seesaw mechanisms of neutrino masses at the TeV scale,''
  Phys.\ Lett.\  B {\bf 679}, 249 (2009)
  [arXiv:0906.1757 [hep-ph]];
  %%CITATION = PHLTA,B679,249;%%
%\cite{Picek:2009is}
%\bibitem{Picek:2009is}
  I.~Picek and B.~Radovcic,
  %``Novel TeV-scale seesaw mechanism with Dirac mediators,''
  Phys.\ Lett.\  B {\bf 687}, 338 (2010)
  [arXiv:0911.1374 [hep-ph]].
  %%CITATION = PHLTA,B687,338;%%

%\cite{Bonnet:2009ej}
\bibitem{Bonnet:2009ej}
  F.~Bonnet, D.~Hernandez, T.~Ota and W.~Winter,
  %``Neutrino masses from higher than d=5 effective operators,''
  JHEP {\bf 0910}, 076 (2009)
  [arXiv:0907.3143 [hep-ph]].
  %%CITATION = JHEPA,0910,076;%%

%\cite{Liao:2010ku}
\bibitem{Liao:2010ku}
  Y.~Liao,
  %``Unique Neutrino Mass Operator at any Mass Dimension,''
  Phys.\ Lett.\  B {\bf 694}, 346 (2011)
  [arXiv:1009.1692 [hep-ph]].
  %%CITATION = PHLTA,B694,346;%%

%\cite{Babu:2001ex}
\bibitem{Babu:2001ex}
  K.~S.~Babu and C.~N.~Leung,
  %``Classification of effective neutrino mass operators,''
  Nucl.\ Phys.\  B {\bf 619}, 667 (2001)
  [arXiv:hep-ph/0106054].
  %%CITATION = NUPHA,B619,667;%%

%\cite{de Gouvea:2007xp}
\bibitem{de Gouvea:2007xp}
  A.~de Gouvea and J.~Jenkins,
  %``A Survey of Lepton Number Violation Via Effective Operators,''
  Phys.\ Rev.\  D {\bf 77}, 013008 (2008)
  [arXiv:0708.1344 [hep-ph]].
  %%CITATION = PHRVA,D77,013008;%%

%\cite{Kayser:1982br}
\bibitem{Kayser:1982br}
  B.~Kayser,
  %``Majorana Neutrinos And Their Electromagnetic Properties,''
  Phys.\ Rev.\  D {\bf 26}, 1662 (1982).
  %%CITATION = PHRVA,D26,1662;%%

%\cite{Auerbach:2001wg}
\bibitem{Auerbach:2001wg}
  L.~B.~Auerbach {\it et al.}  [LSND Collaboration],
  %``Measurement of electron-neutrino electron elastic scattering,''
  Phys.\ Rev.\  D {\bf 63}, 112001 (2001)
  [arXiv:hep-ex/0101039];
  %%CITATION = PHRVA,D63,112001;%%
%\cite{Schwienhorst:2001sj}
%\bibitem{Schwienhorst:2001sj}
  R.~Schwienhorst {\it et al.}  [DONUT Collaboration],
  %``A New Upper Limit for the Tau-Neutrino Magnetic Moment,''
  Phys.\ Lett.\  B {\bf 513}, 23 (2001)
  [arXiv:hep-ex/0102026];
  %%CITATION = PHLTA,B513,23;%%
%\cite{Daraktchieva:2005kn}
%\bibitem{Daraktchieva:2005kn}
  Z.~Daraktchieva {\it et al.}  [MUNU Collaboration],
  %``Final results on the neutrino magnetic moment from the MUNU experiment,''
  Phys.\ Lett.\  B {\bf 615}, 153 (2005)
  [arXiv:hep-ex/0502037].
  %%CITATION = PHLTA,B615,153;%%

%\cite{Raffelt:1999gv}
\bibitem{Raffelt:1999gv}
  G.~G.~Raffelt,
  %``Limits on neutrino electromagnetic properties: An update,''
  Phys.\ Rept.\  {\bf 320}, 319 (1999).
  %%CITATION = PRPLC,320,319;%%

%\cite{KlapdorKleingrothaus:2004ge}
\bibitem{KlapdorKleingrothaus:2004ge}
  H.~V.~Klapdor-Kleingrothaus, A.~Dietz, I.~V.~Krivosheina and O.~Chkvorets,
  %``Data Acquisition and Analysis of the 76Ge Double Beta Experiment in Gran
  %Sasso 1990-2003,''
  Nucl.\ Instrum.\ Meth.\  A {\bf 522}, 371 (2004)
  [arXiv:hep-ph/0403018];
  %%CITATION = NUIMA,A522,371;%%
%\cite{Arnaboldi:2005cg}
%\bibitem{Arnaboldi:2005cg}
  C.~Arnaboldi {\it et al.},
  %``A New Limit on the Neutrinoless DBD of 130Te,''
  Phys.\ Rev.\ Lett.\  {\bf 95}, 142501 (2005)
  [arXiv:hep-ex/0501034].
  %%CITATION = PRLTA,95,142501;%%



%\end{enumerate}
\end{thebibliography}
\end{document}